\documentclass{aa} 

\usepackage{graphicx}
\usepackage{txfonts}
\usepackage{gensymb} 
\usepackage{natbib}
\bibpunct{(}{)}{;}{a}{}{,} 
 
\usepackage{hyperref} 
\hypersetup{colorlinks=true, linkcolor=blue, citecolor=blue, urlcolor=blue}

\newcommand{\sophi}{SO/PHI}
\newcommand{\hmi}{SDO/HMI}
\newcommand{\aia}{SDO/AIA}

\newcommand{\hrt}{SO/PHI-HRT}
\newcommand{\blos}{$B_{\mathrm{LOS}}$}
\newcommand{\eui}{EUI}

\newcommand{\euv}{$\mathrm{HRI_{EUV}}$}
\newcommand{\lya}{$\mathrm{HRI_{Lya}}$}

\newcommand{\orcid}[1]{\protect\href{https://orcid.org/#1}{\protect\includegraphics[width=8pt]{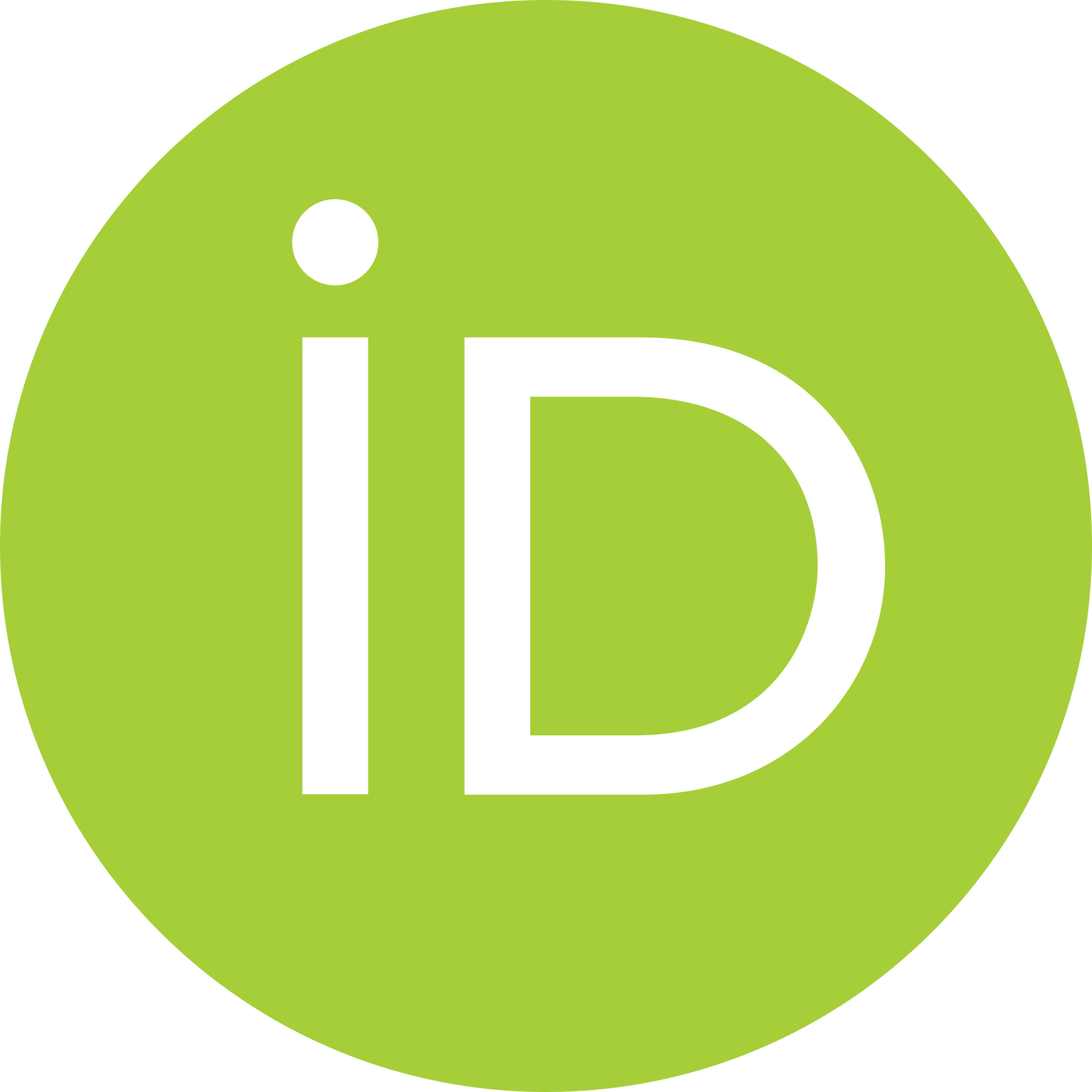}}}

\begin{document}

\title{Magnetic structure of coronal dark halos}
   
\author{J.D.~N\" olke\inst{\ref{i:mps}}\thanks{\hbox{Corresponding author: J.D.~N\" olke} \hbox{\email{noelke@mps.mpg.de}}}\orcid{0000-0003-1988-4494} \and
S.K.~Solanki\inst{\ref{i:mps}}\orcid{0000-0002-3418-8449} \and
J.~Hirzberger\inst{\ref{i:mps}} \and
H.~Peter\inst{\ref{i:mps},\ref{i:kis}}\orcid{0000-0001-9921-0937} \and
L.P.~Chitta\inst{\ref{i:mps}}\orcid{0000-0002-9270-6785} \and
K.H.~Glassmeier\inst{\ref{i:mps},\ref{i:tubs}}\orcid{0000-0003-4327-5576} \and
D.~Calchetti\inst{\ref{i:mps}}\orcid{0000-0003-2755-5295} \and
G.~Valori\inst{\ref{i:mps}}\orcid{0000-0001-7809-0067}
}

\institute{
    Max-Planck-Institut f\"ur Sonnensystemforschung, Justus-von-Liebig-Weg 3, 37077 G\"ottingen, Germany \\ \email{solanki@mps.mpg.de}\label{i:mps}
    \and
    Institut f\"ur Sonnenphysik (KIS), Georges-K\"ohler-Allee 401A, 79110 Freiburg, Germany\label{i:kis}
    \and
    Institut für Geophysik und extraterrestrische Physik, Technische Universität Braunschweig, Braunschweig, Germany\label{i:tubs}
}

   \date{Submitted 27 August 2025, accepted 14 March 2026}

   \abstract
    {At low coronal temperatures around or below 1\,MK distinct areas in the surroundings of active regions (ARs) show emission at a level significantly below the emission coming from the quiet Sun (QS). These areas are referred to as dark halos, dark canopies, or dark moats.}
    {To better understand the nature of dark halos, we studied the connection between the photospheric magnetic field and coronal emission at different temperatures.}
    {Combining Solar Orbiter data from the high-resolution Polarimetric and Helioseismic Imager (\sophi) and Extreme Ultraviolet Imager (\eui) instruments allowed us to identify the areas that are dark in the extreme ultraviolet (EUV) in the immediate vicinity of an AR. We probed  the photospheric magnetic field as well as the coronal intensities as a function of distance to the AR NOAA 12893.}
    {The dark halo has an unsigned magnetic flux density similar to the QS, but shows a strong radial dependence with distance from the AR centre. It drops by 38\% from 6.1\,G at the inner boundary to 3.8\,G at the outer, shifting from above to below QS levels. Coronal emission $\leq$1\,MK is $\sim$40\% below QS and shows no dependence on distance to the AR centre. In contrast, at $\geq$1.6\,MK, emission exceeds QS levels, but declines outwards towards QS values. A few hot loops extend from the AR periphery across the halo, while at lower temperatures no such loops appear and short loops dominate the corona.}
    {The reduced unsigned magnetic flux density in the outermost parts of the dark halo, below QS level, suggests that reduced coronal heating due to weak underlying magnetic flux heating could be partially responsible for the reduced emission around 1\,MK. Closer to the AR, other mechanisms might lead to reduced heating. The different loop structures detected for hotter and cooler coronal temperatures likely play a crucial role in understanding coronal dark halos.}


\maketitle
\nolinenumbers

\section{Introduction}\label{S:introduction}
At low coronal temperatures around or below 1\,MK, active regions (ARs) are often surrounded by extended areas of significantly reduced emission  compared to the emission coming from the quiet Sun (QS). This can be very prominently seen in data obtained by the Atmospheric Imaging Assembly \citep[AIA;][]{SDO/AIA} on board the Solar Dynamics Observatory \citep[SDO;][]{Pesnell2012} at 171\,\AA\ or by the High Resolution Imager (\euv) in the extreme-ultraviolet (EUV) of the Extreme Ultraviolet Imager \citep[EUI;][]{EUI_instrument} on board Solar Orbiter \citep[][]{mueller:2020}. These areas of reduced emission have been referred to by different names: dark canopies \citep{wang2011}, emerging dimmings \citep{zhang2012}, dark halos \citep{Andretta&DelZanna2014}, or dark moats \citep{singh2021}. For our work we have adopted the term dark halos. Dark halos are visible at all transition region temperatures and at low coronal temperatures. However, unlike coronal holes (CHs) they cannot be distinguished from their surroundings at temperatures above 1\,MK.

Several models have been proposed to explain the nature of the coronal dark halos. The model put forward by \citet{wang2011} suggests that dark halos show less emission, due to the presence of extreme-ultraviolet (EUV) absorbing material (i.e. neutral hydrogen or helium). However, as already shown by \citet{Andretta&DelZanna2014} and pointed out by \citet{Lezzi2023}, the dark halos also appear darker at wavelengths longer than the edge of the Lyman continuum at 912\,\AA.

\citet{singh2021} conclude that the formation of the coronal dark halos results from a specific configuration of the magnetic field. The strong magnetic field from the AR reaching over the area of the dark halos presses the underlying magnetic loops down, and thus restricts them in height. According to \citet{Antiochos&Noci1986}, loops below 5\,Mm can only reach temperatures up to $10^5$\,K, and consequently do not emit in the EUV around 1\,MK. As a result, the bulk of this EUV emission is missing inside the dark halos. At $10^5$\,K the dark halos are expected to show strong emission in the AIA 304\,\AA\ channel. However, \citet{singh2021} observed the opposite and stated that further research is required.
Furthermore, modern 3D models produce an abundance of (transient) short loops that reach temperatures well above $10^5$~K \citep[e.g.][]{Chen_etal_2021, Chen_etal_2025} invalidating the argument about missing cool loops based on old model considerations.

In their most recent study, \cite{Lezzi_etal_2024} have reported a fine structure within the dark halos consisting of interconnected EUV-bright bundles and dark regions. They found this fine structure also to be present in transition region lines and in Lyman-alpha. Further, its signature can be seen in magnetograms. From this, the authors conclude that coronal dark halos could have their origin in the lower solar atmosphere.

\citet{zhang2012} and \citet{Payne_etal_2021} observed emerging dimmings, a reduction in emission around ARs that is similar in appearance to dark halos. On the other hand, these emerging dimmings only last for several hours and it is thus not clear whether coronal dark halos and emerging dimmings are the same phenomenon. \citet{zhang2012} explain the reduced emission at lower coronal temperatures as a result of reconnection of the existing flux with the emerging flux. This reconfiguration of the magnetic field in the corona causes the plasma to heat up, thus leading to an increase in emission at higher temperatures and a decrease in the plasma density at lower temperatures.

Chromospheric counterparts of coronal dark halos were first identified in the early 20th century by \cite{Hale+Ellerman_1903} and \cite{StJohn_1911_chromospheric_dark_halos}. Later, \cite{Bumba+Howard_1965} linked these chromospheric structures to dark H$\alpha$ fibrils. However, it remains unclear whether or how they are connected to the coronal dark halos.

Coronal dark halos are only one type of EUV-dark structures in the solar corona. Other known EUV-dark structures are CHs, AR outflows, and coronal voids. CHs \citep{Waldmeier1956,Waldmeier1957} are structured by open magnetic fields and exhibit a significant flux imbalance, while the unsigned magnetic field is similar to the QS \citep{Wiegelmann2004}. 
They appear dark in the EUV because due to the open field the coronal plasma is not within the corona and more energy is placed into acceleration than into heating the plasma, in contrast to ARs or the QS. Consequently, CHs are less hot and less dense \citep{Munro1972,Cranmer2009}. Thus, they can best be seen as dark regions in EUV lines that form at hotter coronal temperatures. During solar activity minimum large persistent CHs are located at the poles. Around solar maximum CHs also show up at lower latitudes. They are referred to as low-latitude, equatorial, or on-disc CHs, and they often appear in the vicinity of magnetically complex ARs.

Active region outflows are characterised by a strong blueshift in a Doppler velocity map mostly at the edge of an AR \citep{doschek2007, doschek2008, sakao2007_ARoutflows, harra2008_ARoutflows}. These AR outflows typically last at least a day, and are rooted in or close to strong unipolar magnetic flux regions, coinciding with quasi-separatrix layers where the magnetic connectivity undergoes rapid changes. Additionally, they are associated with areas of reduced coronal emission due to the reduced density. Some of these outflows follow open magnetic field lines that extend into the heliosphere, while others are at the footpoints of large-scale closed loops \citep[see e.g.][]{baker2009_ARoutflows}.

Coronal voids \citep{Noelke2023} on the other hand are a QS phenomenon. These structures form above areas in the photosphere that are characterised by the absence of stronger magnetic elements. These reduced photospheric magnetic fields lead to a locally reduced heating of the solar corona and consequently to a reduced emission in the EUV.

Our study focused on examining the magnetic field and coronal emission of the dark halos, specifically exploring their correlation with distance from the AR centre. Our aim was to understand how the magnetic field inside the dark halos is structured and how the large-scale field of the AR facilitates the formation of these dark halos. Additionally, we compared our findings on the dark halos with those of CHs and coronal voids. 

\section{Observations}
A cross-calibration campaign for Solar Orbiter with observatories in Earth orbit was performed during the mission's cruise phase on 2021 November 5. At this time Solar Orbiter was almost along the Sun--Earth line, with an angle of only $0.57\degree$ and the distance to the Sun measured 0.86\,au. 
Solar Orbiter was pointing at solar disc centre. The High Resolution Telescope \citep[HRT;][]{gandorfer18} of the Polarimetric and Helioseismic Imager \citep[SO/PHI;][]{PHI_instrument} and the EUI instrument observed at several times during this day. In this study we consider the periods that contain co-temporal observations from both instruments. Because of the advantageous orbit position we complemented the Solar Orbiter data with observations from \aia.

\subsection{Photospheric magnetic field observed with \sophi}
The \hrt\ instrument observed between 22:46 and 23:34 (here and in the following all times are given in UTC) with a cadence of 2.5~minutes. In this study we used a single magnetogram. The \sophi\ spectral scan to obtain this magnetogram was started at 23:29:35 and its total duration was 66~seconds. At the observing distance of 0.86\,au \hrt's plate scale of 0.5\arcsec\ corresponds to 311\,km/pixel on the solar surface. The selected line-of-sight (LOS) magnetogram is a level~2 data product (available on request: sophi\_support@mps.mpg.de) and shows the full field of view (FOV) of $2048\times2048$~pixels. 

\subsection{Coronal observations with EUI and \aia} 
The \eui\ high-resolution telescope in the EUV (\euv) observed between 22:58 and 23:59 with a cadence of 5~seconds. We chose a snapshot at 23:30:00 with an exposure time of 2.8~s that was taken around mid-observation of the \hrt\ magnetogram. Like \hrt\ the \euv\ records $2048\times2048$~pixels. With its slightly different plate scale of 0.492\arcsec\ each pixel corresponds to 307\,km/pixel on the Sun at the observing distance. The \euv\ FOV, therefore, covered $629\times629$\,Mm. We used level~2 data.\footnote{\url{https://doi.org/10.24414/z818-4163}}

The observation time at 23:30:00 on board Solar Orbiter corresponds to an observation time on Earth of 23:31:07 due to the different light travel times. Hence, for our analysis we selected observations taken at this time from three different \aia\ passbands: 171\,\AA, 193\,\AA, and 211\,\AA. For the 171\,\AA\ and the 193\,\AA\ channels observations at 23:31:09 and at 23:31:04, respectively, are used. To reduce the noise level in the 211\,\AA\ channel we averaged five images, spanning one minute, around 23:31 (from 23:30:31 to 23:31:21). 
To study structural details we produced contrast enhanced images. Creating a contrast enhanced image requires an image with a very high signal-to-noise ratio (S/N). This was achieved by averaging a time series of 30~minutes ($\sim$180 images) for each of the three channels. Specifically, we averaged data around 23:31 that had been corrected for solar rotation. 
For all \aia\ observations we produced level 1.5 data and then additionally normalised the level 1.5 data to the exposure time. Therefore, intensities for each pixel are given in the same unit DN/s as the level~2 \euv\ intensities.

\subsection{Spatial alignment of the different datasets}
Since \eui\ has different telescopes for each of the two high-resolution channels in the EUV (\euv) and in the Lyman-alpha line (\lya) it can obtain simultaneous observations for the different channels. As described in \citet{Noelke2023} and \citet{Kahil2022} we aligned the \hrt\ LOS magnetogram to \lya\ because the magnetic or chromospheric network is visible in both, and then we aligned \lya\ with \euv. Unlike in \citet{Noelke2023} and \citet{Kahil2022} we did not apply a geometric distortion correction to the LOS magnetic field (\blos) map because it did not noticeably improve the data quality. To remove dead pixels on the HRT detector a binary mask was applied to the \blos\ map and we additionally removed faulty pixels within two small squares at the right and left upper boundaries of the FOV. In the alignment process all data were re-scaled to the \euv\ plate scale and later cropped to the \euv\ FOV. As a result of this alignment, the \blos\ map needed to be shifted in the south-east direction and hence no magnetic field data is available in the upper part of the \euv\ FOV.

To align the \aia\ data with the \euv\ observation, we first re-projected the \aia\ observations onto the \euv\ frame using the SunPy routine \citep{sunpy_community2020}. The rotation of the \euv\ image against the solar north and hence also between the \euv\ and \aia\ 171\,\AA\ image is $21.5\degree$. Next, we correlated the re-projected 171\,\AA\ image with the \euv\ image and corrected all re-projected data for the small remaining shift between the re-projected 171\,\AA\ and the \euv\ images. 
 
\section{Methods}

\begin{figure*}
\centering
\includegraphics[width=\textwidth]{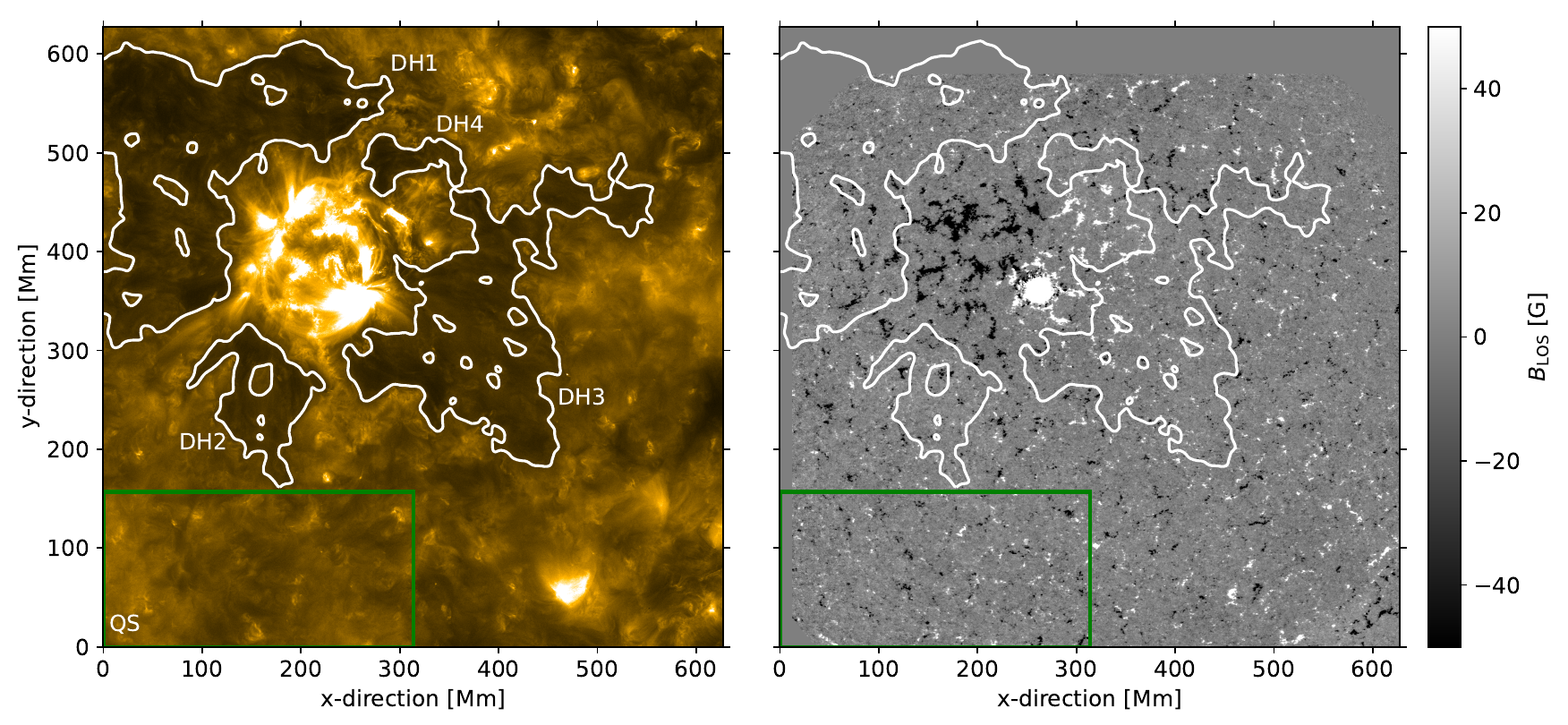}
    \caption{\euv\ image (left) and \hrt magnetogram (right). The 75\% intensity contours of the dark halo are plotted in white on the \euv\ image and are projected onto the \hrt magnetogram. The individual dark halo patches are labelled in white on the \euv\ image. The green rectangle at the bottom indicates the area of the reference QS.}
    \label{overview_EUI+PHI}
\end{figure*}

\subsection{Defining the dark halo boundaries}\label{S:dark_halo_boundaries}
The Solar Orbiter \euv\ channel observed an area of $629\times629$\,Mm at disc centre (see Fig.~\ref{overview_EUI+PHI}). Within this FOV the AR NOAA 12893 can be seen in the upper part, while the lower part shows a QS area. We defined the area of the left bottom quarter below $y=157$\,Mm and left of $x=314$\,Mm as our reference QS. 
We chose this area because it is the largest QS area without dark halo patches within the FOV. We further sought to ensure that we did not include any non-QS structures: on the right an on-disc CH is present (see Fig.~\ref{AIA_images}) and the bottom right contains a strong bipolar region which harbours a small pore.

We defined the threshold for the dark halo the same way as the threshold for coronal voids \citep[][]{Noelke2023}: 75\% of the mean QS intensity. 
Since we selected a QS area that exhibits a roughly symmetrical distribution of its EUV-intensities, the mean and median value do not differ greatly.
 
To obtain smooth iso-contours, we convolved the \euv\ image with a Gaussian kernel with a full width at half maximum (FWHM) of 7.2\,Mm (corresponding to a $\sigma$ of 3\,Mm or 10~pixels). We then applied the intensity threshold of 75\% of the mean QS intensity to this convolved image. The dark halo was identified as the EUV-dark areas surrounding the AR. We varied the threshold slightly to test for its robustness and found that our results remain consistent against small changes. Fig.~\ref{overview_EUI+PHI} shows the \euv\ image and \hrt\ magnetogram. The contours of the dark halo patches have been projected onto both images.

\subsection{Determining the AR's magnetic centre and creating ring plots}\label{M:ring_plots}
Since the dark halos form in proximity to ARs it is plausible to assume that a dark halo is not uniform throughout its entire structure. Therefore, to see how the properties of the dark halo are correlated with distance to the AR, we derived these values inside concentric rings around the AR.

First, we determined the magnetic centre of the AR which is used as the centre of the rings. Here we used the geometric centre or centroid which is calculated as the average position of the pixels, treating each pixel equally by considering only their coordinates and not the pixel values. This is achieved by calculating the geometric centre of all pixels with a field strength $|B|$ of at least 100\,G within 215\,Mm (700~pixels) around the AR.
We started at the clearly visible leading sunspot and iterated several times until the derived centre did not change any more. 

We examined different definitions of the AR centre, including a purely geometric centroid and centres weighted by magnetic flux density using various flux-density thresholds. The leading positive polarity is concentrated in a strong sunspot, whereas the opposite polarity is weaker and more spatially dispersed. As a result, weighted or high-threshold methods place the AR centre close to the leading sunspot, while the geometric centre with a relatively low threshold lies near the visual centre of the AR. We also determined a centre based on the brightest EUV emission, which yields a location comparable to that derived from the magnetic field. The resulting radial intensity and unsigned magnetic flux profiles remain qualitatively the same for all tested centre definitions that place the centre near the visual middle of the AR, indicating that the results are robust against reasonable variations in the choice of AR centre. We therefore adopted the geometric centre, which provides a consistent and generally applicable definition.

\begin{figure*}
\centering
\includegraphics[width=\textwidth]{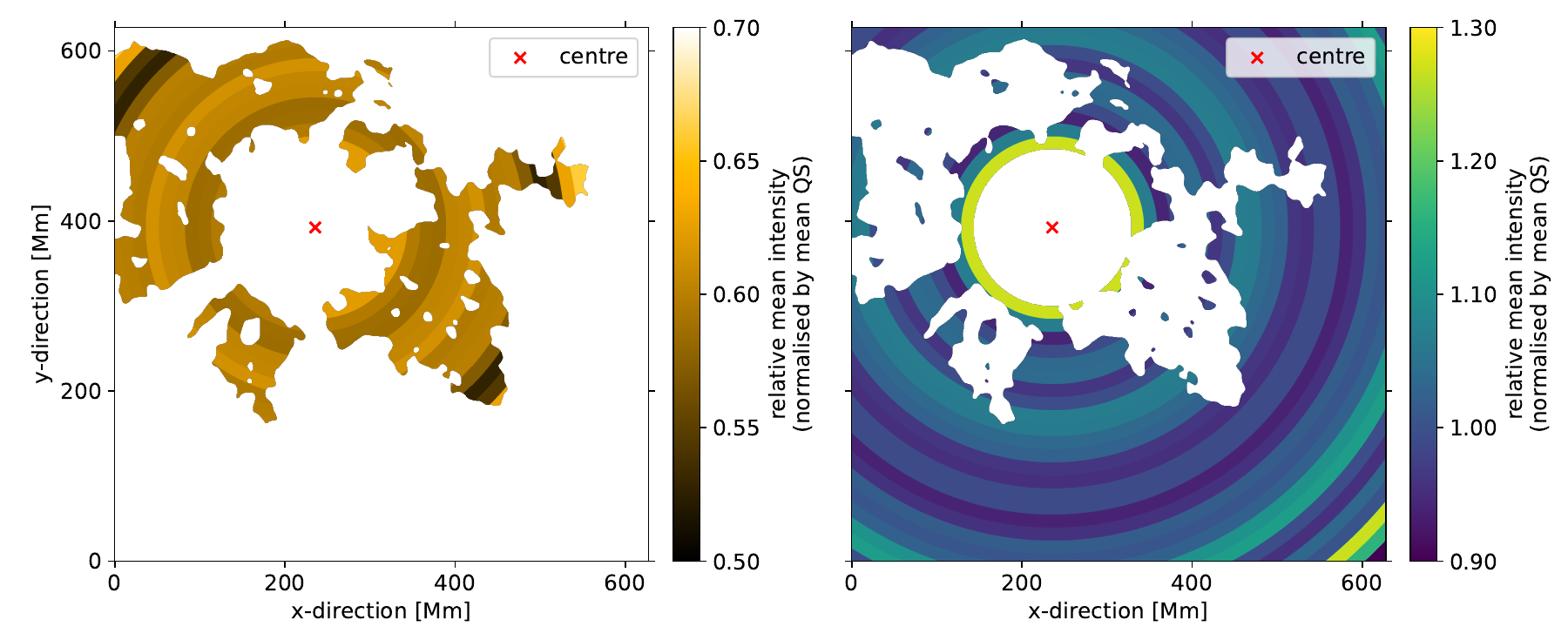}
    \caption{Relative mean \euv intensity inside rings around the AR obtained by dividing the mean intensity in a ring by the mean value of the reference QS. The left and right panels show respectively the relative mean intensity inside the dark halo (i.e. averaged over all dark halo patches at a certain distance to the AR and not treated separately) and for the brighter areas in between the individual dark halo patches.}
    \label{EUI_intensity_rings}
\end{figure*}

Next, concentric rings were drawn around this centre starting roughly at the inner boundaries of the dark halo at 92\,Mm (300~pixels) distance to this centre. The rings have a width of 15\,Mm (50~pixels) each on the \euv\ image to which we geometrically re-scaled all other images.
An example of the EUV brightness in the set of rings can be seen in Fig.~\ref{EUI_intensity_rings} for the average intensities of the \euv\ image. Within each ring the average intensity or unsigned magnetic flux density was calculated either purely as a function of distance, not differentiating between individual dark halo patches; for an individual dark halo patch; or for the brighter areas in between the dark halo patches. For the dark halo the innermost and outermost rings are slightly larger to fully encompass the inner and outer edges of the dark halo.

\section{Results} 
In this section we studied the coronal emission and unsigned magnetic flux density of the coronal dark halo and the brighter areas in between the dark halo patches, as well as individual dark halo patches including a dark halo patch which overlaps with an equatorial CH. We determined how these properties behave as a function of distance to the AR centre.

\begin{figure*}
\centering
\includegraphics[width=\textwidth]{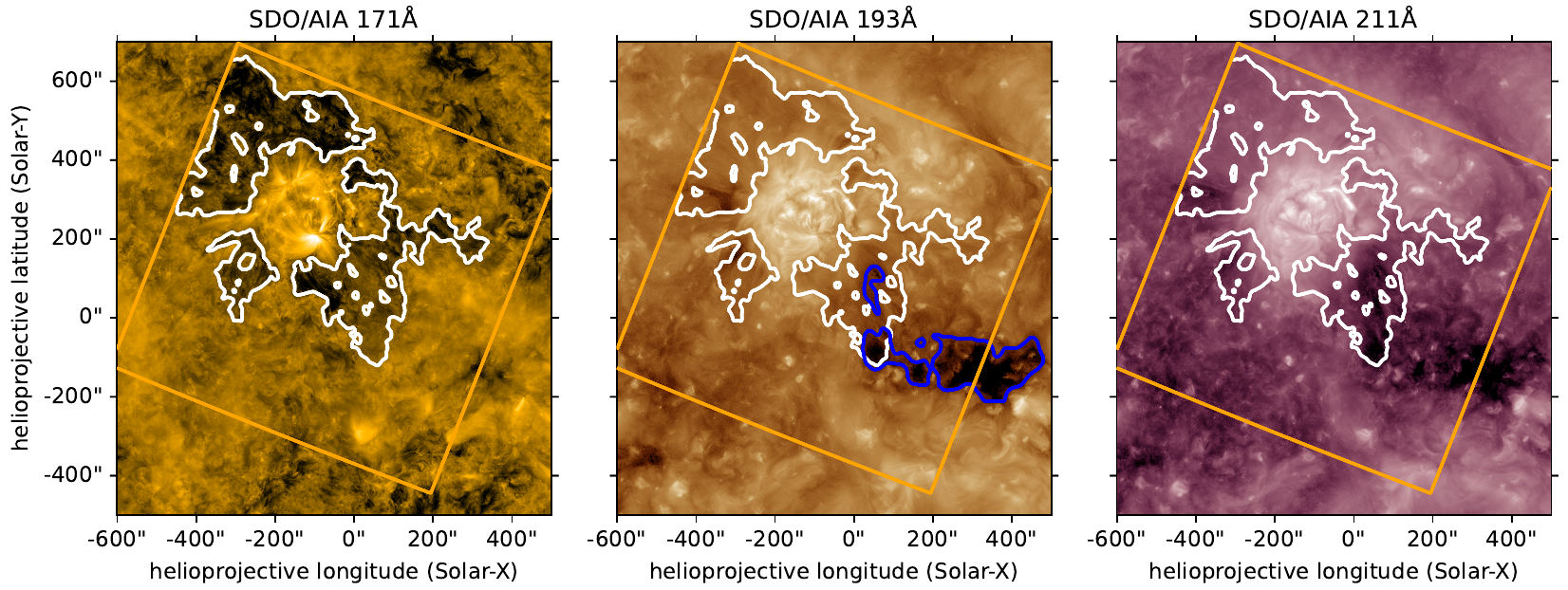}
    \caption{EUV emissions for three different \aia\ passbands: 171\,\AA\ (left), 193\,\AA\ (middle), and 211\,\AA\ (right). The contours of the dark halo (white, as in Fig.~\ref{overview_EUI+PHI}) have been projected onto each of the images; the orange square shows the FOV of the \euv\ observation. In the 193\,\AA\ image we also show the extent of the equatorial CH (blue contour) derived by an intensity threshold of 40\% of the channels mean disc intensity.}
    \label{AIA_images}
\end{figure*}

\begin{figure}
\centering
    \resizebox{\hsize}{!}{\includegraphics{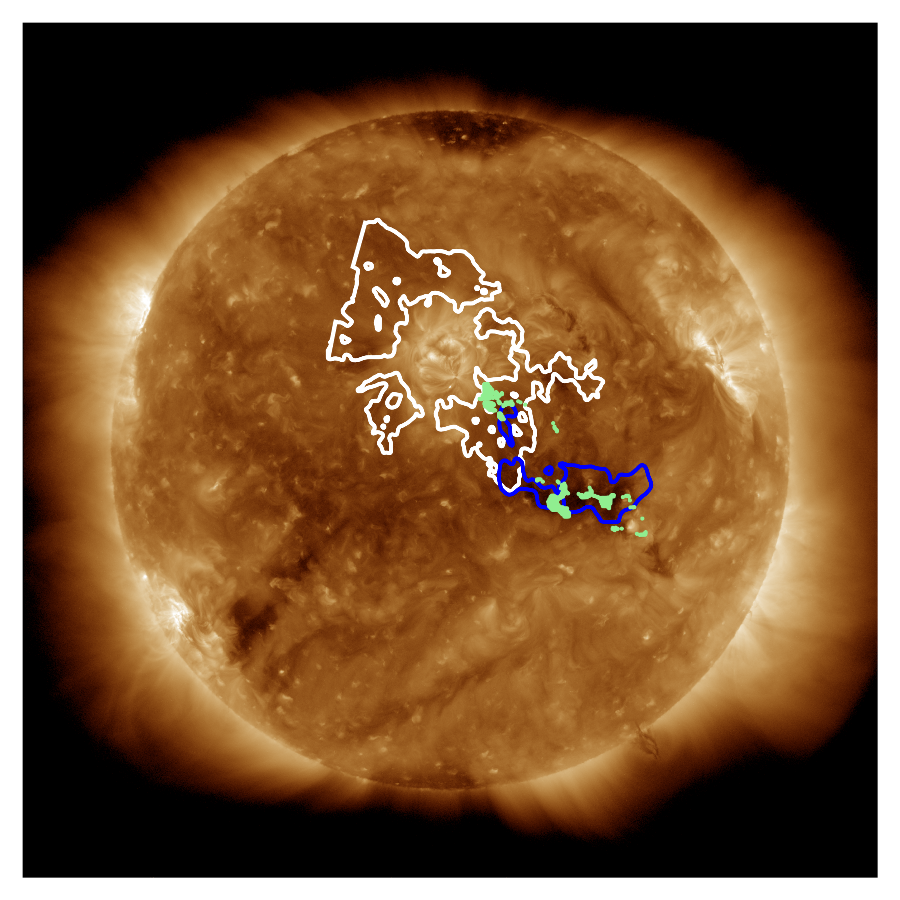}}
    \caption{Open magnetic field originating from the CH adjacent to the AR (blue contours), including from within the dark halo patch DH3 (white contours). The image shows the \aia\ 193\,\AA\ channel on 2021 November 6 12:00 with the footpoints of open field lines in green.}
    \label{open_field}
\end{figure}

\begin{figure}
\centering
    \resizebox{\hsize}{!}{\includegraphics{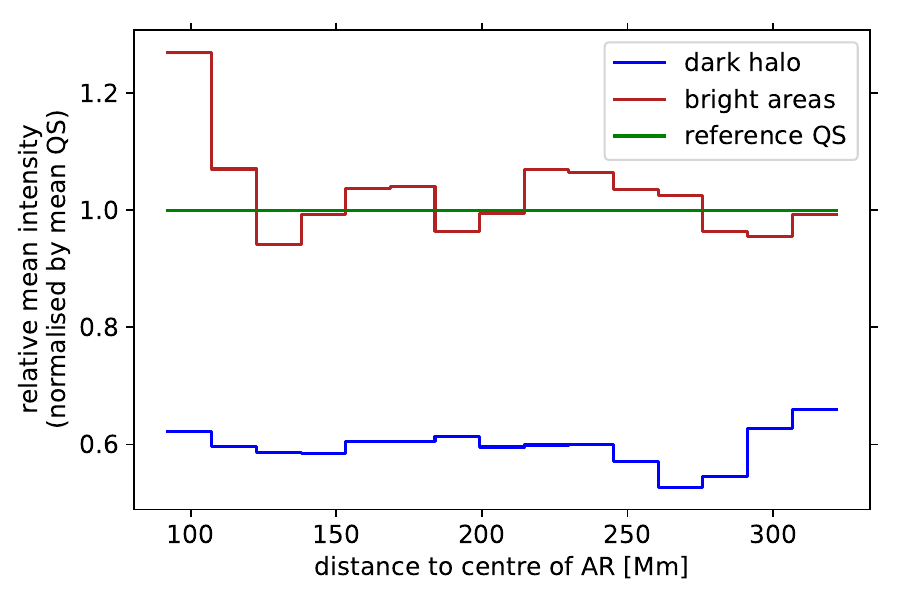}}
    \caption{Relative mean \euv intensity as a function of distance to the AR. The mean intensity has been normalised by the mean value derived for the reference QS. The relative mean intensity inside the dark halo (blue) and for the brighter areas in between the dark halo patches (dark red) are shown. The green line denotes the value for the reference QS.}
    \label{EUI_intensity_distance}
\end{figure}

\subsection{Dark halo and CH overlap}\label{S:dark_halos_and_CH}
The AR is surrounded by four dark halo patches, two larger ones and two smaller in size. We labelled the different dark halo patches DH1, DH2, DH3, and DH4 (see Fig.~\ref{overview_EUI+PHI} for the labels). One of these larger dark halo patches, DH3, is a complex structure since it partially overlaps with a CH. 
The boundaries of the CH were identified via an intensity threshold of 40\% of the median disc intensity in the \aia\ 193\,\AA\ channel. This CH is adjacent to the AR and a smaller separate part of the CH overlaps with a dark halo patch. The contours of both the CH and the dark halo are displayed in Fig.~\ref{AIA_images}. 
Subsequently, to confirm whether the structure is indeed a CH, we checked for the presence of open magnetic fields. To achieve this, we conducted magnetic field extrapolations using pfsspy \citep{Stansby2020_pfsspy,pfsspyNumericalMethods} on the Daily Update Synoptic Frames map of 2021 November 6 taken with the Helioseismic and Magnetic Imager \citep[HMI;][]{Schou2012_HMI} on board SDO, and placed seeds in the areas covered by the dark halo patch DH3 and the larger CH. We then traced the field lines from these seeds at a height of 2\,Mm above the photosphere and set the source surface to 2.5 solar radii. This way we could determine that any open field originates from within the area of the CH and also from the dark halo--CH complex DH3. Fig.~\ref{open_field} displays the open field lines in green on the \aia\ 193\,\AA\ image. 

\subsection{Coronal emission at different temperatures}\label{S:coronal_emission}
Inside the individual dark halo patches the emission in the \euv\ channel is strongly reduced. In particular, the dark halo patches emit only between 57\% and 64\% of the mean emission coming from the reference QS (see Table~\ref{table} for absolute values).

\begin{figure}
\centering
    \resizebox{\hsize}{!}{\includegraphics{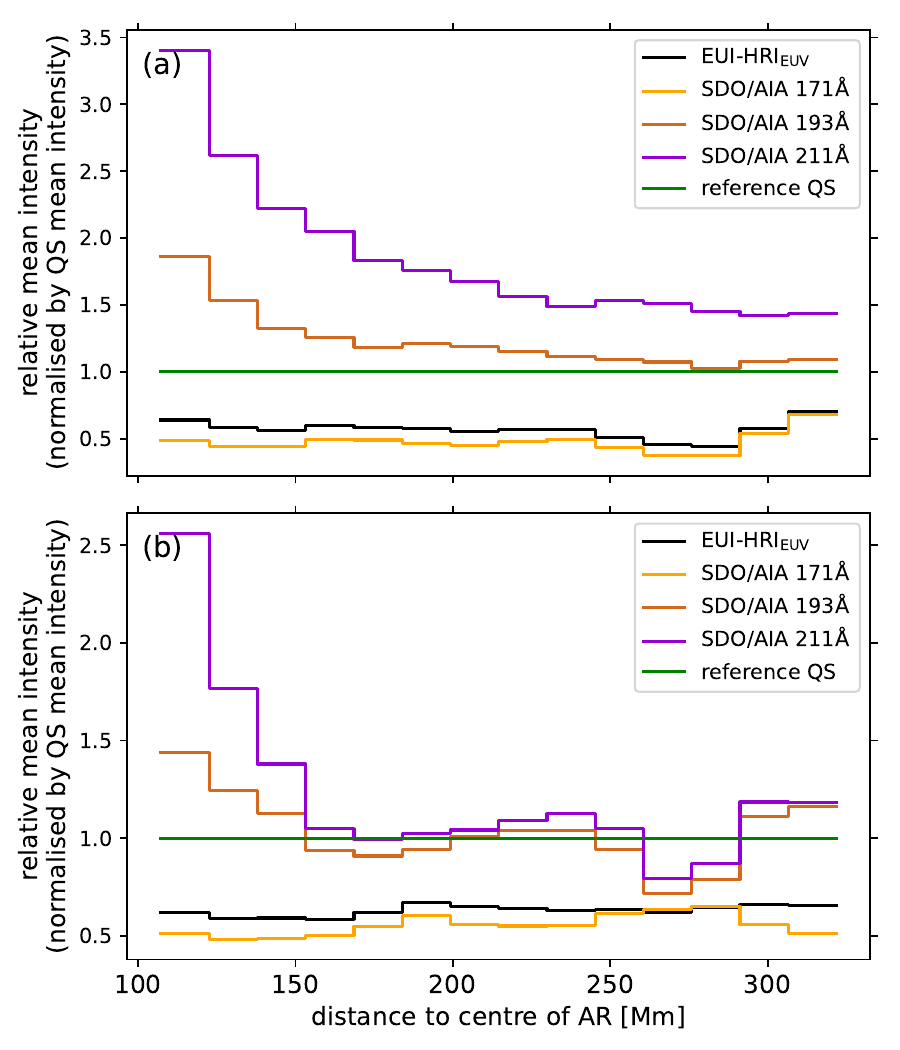}}
    \caption{Relative mean intensities inside individual dark halo patches as a function of distance to the AR shown for \euv\ and the \aia\ 171\,\AA, 193\,\AA, and 211\,\AA\ channels. The mean intensity has been normalised by the mean value of the reference QS of the respective channel. This mean value of the QS is indicated by the green line. Panel~(a) shows the emission from DH1, and panel~(b) from the dark halo--CH complex, DH3.}
    \label{emission_distanceAR_DH1+DH3}
\end{figure}

\begin{figure*}
\centering
\includegraphics[width=\textwidth]{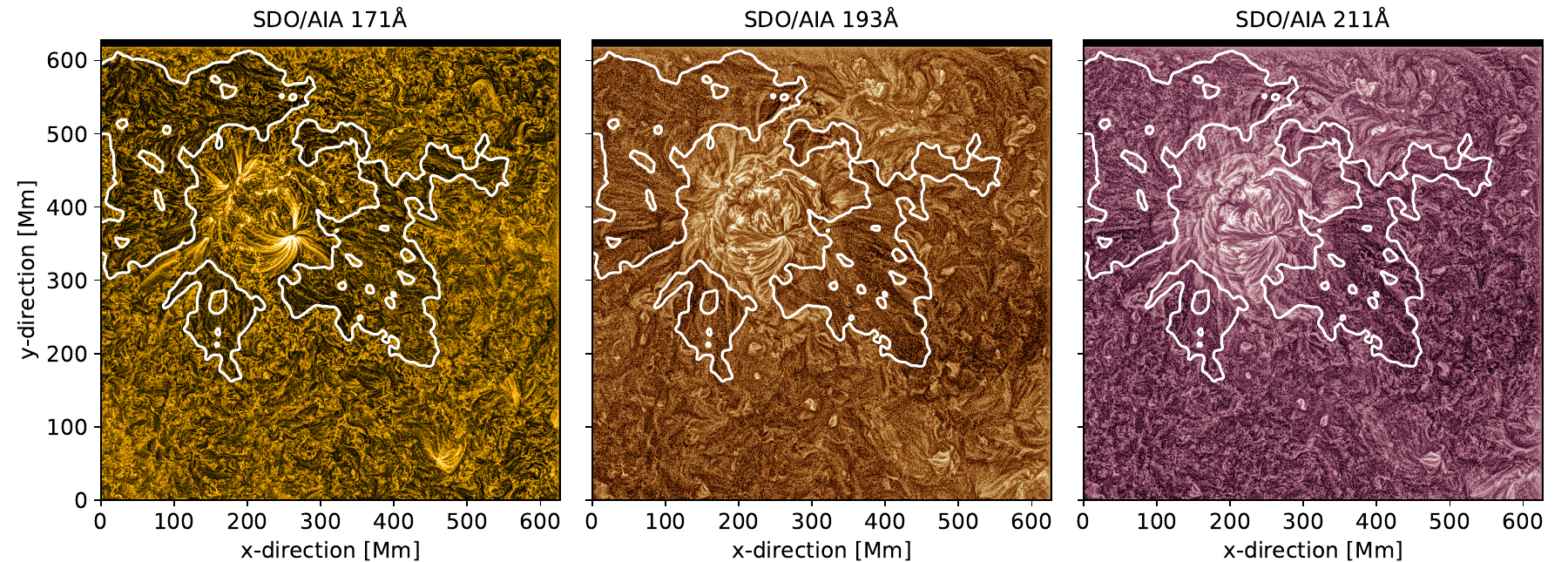}
    \caption{Enhanced images of the three \aia\ passbands 171\,\AA\ (left), 193\,\AA\ (middle), and 211\,\AA\ (right). The white contours outline the dark halo surrounding the AR and the orange square indicates the \euv\ FOV.}
    \label{Enhanced_images}
\end{figure*}

\subsubsection{Mean-intensity's dependence on distance to AR centre}\label{S:intensities_radial_distance}
We computed the relative mean \euv intensity as a function of distance to the AR centre (shown in Fig.~\ref{EUI_intensity_rings}), following the procedure described in Sect.~\ref{M:ring_plots}. Inside each individual ring around the magnetic centre of the AR we derived the relative mean \euv intensity (i.e. the mean intensity inside a ring divided by the mean intensity in the reference QS). In Fig.~\ref{EUI_intensity_rings} we illustrate this for the dark halo and for the bright areas in between the dark halo patches. The area inside each ring is coloured according to the derived relative mean value for the area the ring covers. 
The dark halo and bright areas exhibit very different behaviours. On the one hand, the intensity inside the dark halo stays clearly below QS level and does not vary much. It merely increases slightly towards the inner and outer boundaries. The intensity in the brighter areas, on the other hand, is 1.3 times that of the QS at its inner part. With increasing distance to the AR centre the intensity then drops towards a QS-intensity level. Additionally, the rings on the bottom right, covering a bipolar region, show higher intensities again.

Radial cuts through the two areas displayed in Fig.~\ref{EUI_intensity_rings} are shown in Fig.~\ref{EUI_intensity_distance}. The range over the dark halos is plotted from $92-330$\,Mm distance. For each value we determined the uncertainty of the mean $\sigma_M$ as the standard deviation $\sigma$ divided by the square root of the number of data points $N$, 
\begin{equation}\label{Eq.error_of_mean}
    \sigma_M = \frac{\sigma}{\sqrt{N}} ,
\end{equation}
which is small compared to the mean values measured for the intensities.

The relative mean intensities for the dark halos are clearly much lower and stay below 0.7 at all times. As described above, the relative mean intensities in the brighter region initially drop from a higher than QS values to approximately the QS value and remain relatively constant at greater distances to the AR centre. At all distances there is a difference of at least 0.3 times the QS intensity between the dark halo and the brighter areas in between the individual patches. While this difference is formally a consequence of defining the regions via an intensity threshold, it also reflects the intrinsic darkness of the dark halo patches themselves.

\subsubsection{Coronal emission at different temperatures in individual dark halo patches}\label{S:intensities_individualpatches_radialdistance}
We further wanted to see whether individual dark halo patches behave differently, in particular if DH3, the dark halo--CH complex, shows any significant deviation. Additionally, in order to distinguish between different dark halo models (see Sect.~\ref{S:introduction}), it is essential to understand how the relative intensity behaves at different temperatures. For this we looked at co-temporal \aia\ observations. We chose three different \aia passbands: 171\,\AA, 193\,\AA, and 211\,\AA. The \euv\ and \aia\ 171\,\AA\ channels are very similar yet not identical. While \euv\ has a passband at 174\AA\ and samples the solar corona at 1\,MK with contributions from both the \ion{Fe}{ix} and \ion{Fe}{x} lines \citep[][]{EUI_instrument}, the 171\,\AA\ \aia\ channel has most of its contributions from the \ion{Fe}{ix} line and probes the corona just below 1\,MK \citep[][]{SDO/AIA,Peter_etal_2012}.
The \aia\ 193\,\AA\ and 211\,\AA\ channels see the corona at the higher temperatures of 1.6\,MK, and 2.0\,MK, respectively.

Iso-contours of the individual dark halo patches overlaid onto each \aia image are shown in Fig.~\ref{AIA_images}. Similarly to our approach with the \euv\ data, we calculated the mean intensity within each ring (using the same AR centre and ring width as for the \euv\ data). This allows us to present the relative intensity as a function of distance for the different temperature channels, as shown in Fig.~\ref{emission_distanceAR_DH1+DH3}. 
They are both included in the figure to facilitate a direct comparison with the \euv\ channel and the reference QS. 

We conducted an analysis of the emission coming from the dark halo patch DH1 (see Fig.~\ref{emission_distanceAR_DH1+DH3}a).
The 171\,\AA\ \aia\ channel imaging cooler coronal plasma shows a rather uniformly reduced emission inside the dark halo patch, very similar to the \euv\ channel. At hotter temperatures the emission from the dark halo patch DH1 has a very different behaviour: For the 193\,\AA\ channel the relative intensity at the inner boundary is nearly twice that of the QS in this channel. In the 211\,\AA\ channel the dark halo patch is even brighter and the relative intensity is more than three times as intense as in the QS at 211\,\AA. In both channels the relative mean intensity then decreases with increasing distance to the AR centre. Both channels exhibit emission above the QS level at all distances. At the outer boundary of the dark halo patch the relative intensities in the 193\,\AA\ passband are at about the level as the QS, while the 211\,\AA\ channels still shows more emission from the outer boundaries of the dark halo patch than from the mean QS.

The second dark halo patch for which we analysed the emission at different temperatures is the dark halo-CH complex DH3 (Fig.~\ref{emission_distanceAR_DH1+DH3}b). We aimed to determine whether this structure behaves differently due to its CH contribution. At coronal temperatures around 1\,MK (seen by the \euv\ and 171\,\AA\ channels) DH3 shows a very similar behaviour to DH1. In the channels imaging hotter plasma (193\,\AA\ and 211\,\AA), the general trend also resembles that of DH1, showing mostly emission stronger than the QS level and a decrease towards larger distances from the AR. However, two notable divergences are observed: at distances of 175\,Mm and 280\,Mm from the AR centre the intensities drop to or below the QS level before increasing again. These distances coincide with the areas where the embedded CH is located. While the CH’s presence is distinctly visible in specific locations, the overall behaviour of DH3 is very similar to that of DH1.

\subsubsection{Bright loops over dark halos}\label{S:long_loops}
To better highlight coronal structures in and around the dark halo patches, we applied the method described by \citet{morgen+druckmueller_enhanced_images} which uses a multi-scale Gaussian normalisation to enhance the image contrast. It is implemented in the python packages sunkit\_image.enhance and we used this package to create contrast enhanced images of the three different \aia\ channels (visible in Fig.~\ref{Enhanced_images}). 
In the enhanced image of the \aia\ 171\,\AA\ channel bright loops originating from the periphery of the AR terminate approximately at the inner boundaries of the dark halo. There are no bright loops visible which cross over the dark halo patches or reach deep inside. At this temperature around 0.8\,MK the corona is dominated by small-scale loops. Fewer of these small-scale loops appear within the dark halo than outside. The enhanced images of the passbands probing hotter plasma, on the other hand, show bright loops over the dark halo patches. This difference in loop structure is further highlighted in Appendix~\ref{appendixB}. Similar observations were also made by \cite{singh2021}. A discussion of their and our interpretation of this result can be found in Sect.~\ref{S:discussion_nature}. 
In detail, we note that elongated brighter loops extend radially from the AR. They begin at the periphery of the AR and either cross the dark halo or terminate within it, some ending inside the halo itself or within the brighter islands it encloses. Other bright loops extend from within the dark halo, connecting either to the halo itself or to the bright islands it encloses, while some originate in these regions and stretch outwards beyond the halo’s boundaries. The bright loops are rooted in strong polarity patches located inside the dark halo, the bright islands, and outside the dark halo boundaries. In fact, most of the stronger polarity patches within the dark halo and the bright islands seem to be footpoints of such bright and hot loops.

\subsection{Photospheric magnetic field}\label{S:magnetic_field}
Next, we studied the photospheric magnetic field of the dark halo patches and compared it to the magnetic field of the bright areas between them, and to the QS magnetic field.

\subsubsection{Average unsigned magnetic flux density}\label{S:averageBLOS}

\begin{table*}
\caption{Unsigned magnetic flux density and coronal emission in individual dark halo patches.}     
\label{table}     
\centering                                     
\begin{tabular}{c c c c c c}         
\hline\hline                       
Dark halo patch & $\langle$|\blos|$\rangle$\tablefootmark{b} & \multicolumn{2}{c}{Imbalance in magnetic flux} &  \multicolumn{2}{c}{Emission (\euv)}\\ 
& [G] & $\Delta F$ [G] & rel. $\delta F$ [\%] & abs. [DN/s] & rel. to reference QS [\%]\\ 
\hline                                  
    DH1 & 6.1 & $-1.6$ & 10 & 339 & 57\\     
    DH2 & 4.9 & $+3.0$ & 21 & 381 & 64\\
    DH3 & 4.7 & $+2.2$ & 15 & 373 & 62\\
    DH4 & 5.0 & $+5.2$ & 37 & 355 & 59\\
\hline   
    Entire dark halo\tablefootmark{a} & 5.0 & $+0.6$ & 4 & 357 & 60 \\
    QS & 5.3 & $+0.8$ & 6 & 599 & 100\\
\hline
\end{tabular}
\tablefoot{%
\tablefoottext{a}{Averaged over the whole area covered by the dark halo patches.}
\tablefoottext{b}{Pixels below the 1$\sigma$ noise level of 7.4\,G are set to zero in the computation providing a lower limit for the magnetic flux density.}}
\end{table*}

\begin{figure}
\centering
    \resizebox{\hsize}{!}{\includegraphics{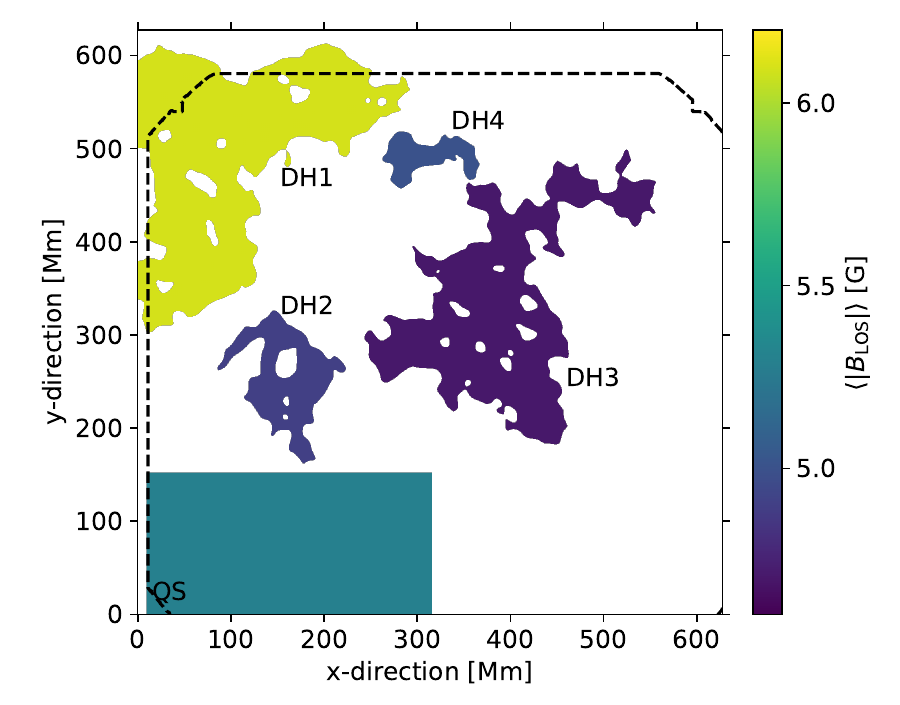}}
    \caption{$\langle$|\blos|$\rangle$ value above the noise level of $1\sigma=7.4$\,G of individual dark halo patches and for the reference QS. The dashed line indicates the \hrt\ FOV.}
    \label{blos_in_individual_dark_halos}
\end{figure}

For each of the dark halo patches we derived the unsigned magnetic flux density, $\langle$|\blos|$\rangle$, individually and compared them to the reference QS.
We derive the unsigned magnetic flux density as the unsigned magnetic flux $\phi_{unsigned}$ per area $A$:
\begin{equation}\label{Eq.fluxdensity}
    \langle |B_{\mathrm{LOS}}| \rangle = \frac{\phi_{unsigned}}{A} = \frac{\int |B_{\mathrm{LOS}}|\, dA}{\int dA}.
\end{equation}
The $\langle$|\blos|$\rangle$ in each dark halo patch was obtained from the LOS magnetogram by first setting pixels below the noise level of $1\sigma=7.4$\,G to zero (see \citealt{sinjan22} for a sensitivity analysis of this SO/PHI-HRT dataset). Next, the mean of the absolute values was calculated. This procedure yields a lower boundary estimate of the real $\langle$|\blos|$\rangle$.
 
The values for the individual dark halo patches are shown in Fig.~\ref{blos_in_individual_dark_halos} and listed in Table~\ref{table}. The $\langle$|\blos|$\rangle$ values inside dark halo patches DH2, DH3, and DH4 are slightly below the 5.3\,G we derived for the reference QS. The patch DH1 lies above this reference QS value. 
However, it is also the only dark halo patch located within the strongly scattered negative polarity, whereas most of the positive polarity is concentrated in the leading sunspot outside the dark halo contours. The inclusion of negative plage areas likely explains the higher $\langle$|\blos|$\rangle$ value in DH1. In addition, a small part of the outer region of DH1 is not covered by the SO/PHI-HRT magnetogram. As shown in Sect. 4.3.2, the magnetic elements are not evenly distributed within the dark halo patch; therefore, sampling only part of the halo could in principle lead to an overestimation of $\langle$|\blos|$\rangle$. However, the area not covered is small compared to the total patch, and the resulting effect is negligible, as we could confirm using \hmi\ data.
The average of all dark halo patches is 5.0\,G. Therefore, the $\langle$|\blos|$\rangle$ inside the dark halo patches is on average very close to that of the QS. 

\subsubsection{Unsigned magnetic flux density's dependence on distance to AR centre}\label{S:BLOS_distance}

\begin{figure}
\centering
    \resizebox{\hsize}{!}{\includegraphics{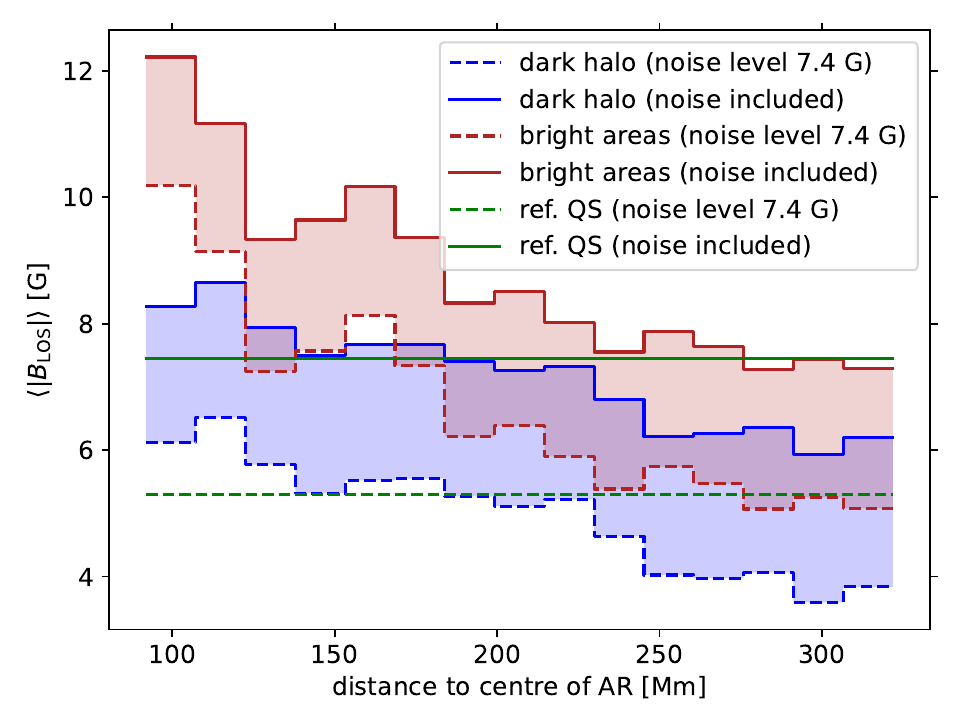}}
    \caption{Unsigned magnetic flux density, $\langle$|\blos|$\rangle$, as a function of distance to the AR centre. The $\langle$|\blos|$\rangle$ inside the dark halo (i.e. averaged over all dark halo patches at a certain distance to the AR centre and not treated separately) is shown in blue; the red curve shows the $\langle$|\blos|$\rangle$ for the bright areas in between the dark halo patches. The dashed lines show the unsigned magnetic flux density calculated with pixels below the noise level of $1\sigma=7.4$\,G set to zero, while the solid lines denote the unsigned flux density including pixels with values below the noise level. The real $\langle$|\blos|$\rangle$ lies within the shaded areas indicated by these upper and lower boundaries. The mean value of the QS is shown in green.}
    \label{blos_distanceAR}
\end{figure}

\begin{figure}
\centering
    \resizebox{\hsize}{!}{\includegraphics{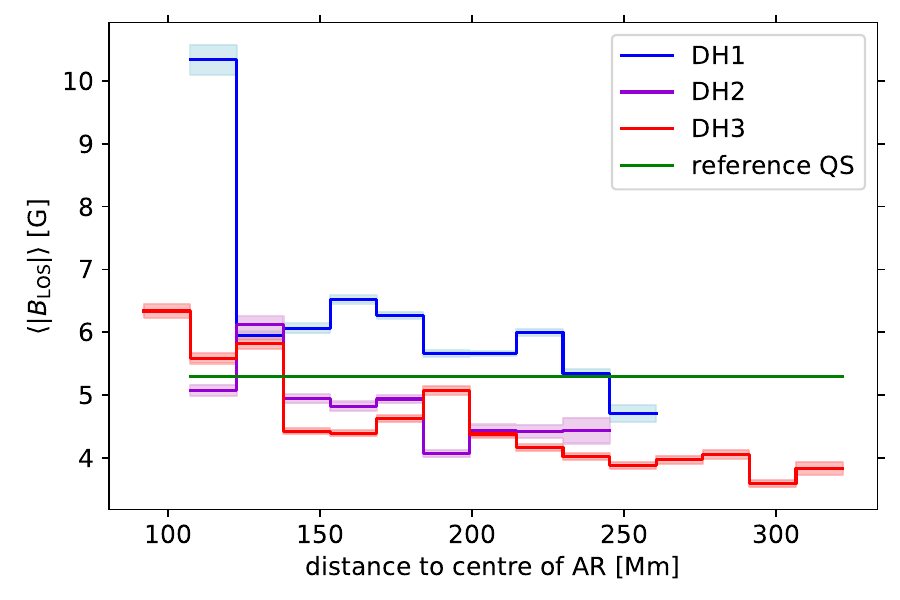}}
    \caption{Unsigned magnetic flux density, $\langle$|\blos|$\rangle$, as a function of distance to the AR centre for individual dark halo patches for a noise level of $1\sigma=7.4$\,G. The values shown represent the lower limit, obtained by setting all pixels below the noise level to zero. The lighter colours denote the uncertainty of the mean $\sigma_M$. The mean value of the QS is shown in green.}
    \label{blos_individualDHs_radially}
\end{figure}

The coronal emission at higher temperatures strongly depends on the distance to the centre of the AR, see Fig.~\ref{emission_distanceAR_DH1+DH3}. Therefore, we also studied the dependence of the underlying magnetic field on distance.
For this, we calculated $\langle$|\blos|$\rangle$ (according to Eq.~\ref{Eq.fluxdensity}) in rings around the AR centre in the same way as we did for the coronal intensities described in Sect.~\ref{S:intensities_radial_distance}. Once again we determined the properties both inside the dark halo and in the brighter areas surrounding them. For both domains separately, we calculated $\langle$|\blos|$\rangle$ in each ring in two different ways to obtain a lower and upper boundary for the real unsigned magnetic flux density: a) within the selected area pixels below the noise level were set to zero and then the $\langle$|\blos|$\rangle$ was calculated like in Sect.~\ref{S:averageBLOS}, and b) pixels below the noise level were also included for calculating the $\langle$|\blos|$\rangle$. In the first case the real unsigned magnetic flux density is underestimated, while the latter overestimates it. Together they provide an upper and lower boundary for the real unsigned magnetic flux density.

For both the dark halo and for the brighter areas in between the individual patches we see a strong decrease away from the AR of the measured $\langle$|\blos|$\rangle$. With and without a threshold on the noise level (i.e. from both the continuous and the dashed set of lines in Fig.~\ref{blos_distanceAR}) the relative trends, and therefore the conclusions that can be drawn, are the same. In particular, referring to the calculation with a noise level applied, Fig.~\ref{blos_distanceAR} shows a decrease in $\langle$|\blos|$\rangle$ inside the dark halo by 38\% from 6.1\,G to 3.8\,G and inside the brighter areas by 50\% from 10.2\,G to 5.1\,G. However, $\langle$|\blos|$\rangle$ inside the dark halo is significantly lower than in the brighter in-between areas (see Fig.~\ref{blos_distanceAR}). Specifically, inside the dark halo $\langle$|\blos|$\rangle$ clearly drops below the QS level at a distance to the AR centre of 230\,Mm and decreases further. In the bright in-between areas $\langle$|\blos|$\rangle$ remains above the QS everywhere. The $\langle$|\blos|$\rangle$ in these areas gradually trends towards the QS level with increasing distance to the AR centre. We further examined $\langle$|\blos|$\rangle$ within the individual dark halo patches DH1, DH2, and DH3 (shown in Fig.~\ref{blos_individualDHs_radially}). The patch DH4 was excluded from the analysis due to its small size. The analysis also focused on potential differences between dark halo patches DH1 and DH2 on the one hand, and DH3 on the other, which includes the CH contribution. We can report that none were found for DH2 and DH3: both dark halo patches exhibit trends consistent with those observed for the averaged dark halo in Fig.~\ref{blos_distanceAR}. The patch DH1, however, deviates from this behaviour. This dark halo patch shows a very strong $\langle$|\blos|$\rangle$ in the innermost ring. The $\langle$|\blos|$\rangle$ barely drops below the QS level, and this occurs further away from the AR centre than in the other patches. Two factors might play a role here: Firstly, the dark halo patch lies within the negative polarity region, which is much more extended than the more concentrated positive polarity. Secondly, this patch is not fully covered by the \hrt. Using SDO/HMI data, we confirmed that, similar to the other patches, it has low magnetic flux density in this area furthest from the AR centre. It is therefore plausible that, for this dark halo patch, $\langle$|\blos|$\rangle$ continues to decrease with increasing distance from the AR centre.
In conclusion, the common behaviour among most individual dark halo patches and the averaged dark halo is that $\langle$|\blos|$\rangle$ consistently drops below the QS level. 

\subsubsection{Flux imbalance in dark halos}\label{S:flux_imbalance}
To check the imbalance in the magnetic flux inside the dark halo patches, similar to \citet{Noelke2023} for coronal voids, we define the flux imbalance as the absolute value of the mean \blos, 
\begin{equation}\label{Eq.abs.imbalance}
    \Delta F = |\left<B_{\rm LOS,t}\right>| ,
\end{equation}
again only including pixels with field strength above the noise level $t$, where the spatial averages are given by $\left< ... \right>$.
Similarly, the relative imbalance in magnetic flux 
\begin{equation}\label{Eq.rel.imbalance}
    \delta F = {|\left< B_{\rm LOS,t}\right> | \over \left< |B_{\rm LOS,t}| \right>} \,
\end{equation}
is defined as the absolute flux imbalance divided by the mean unsigned field strength. 
We derived the flux imbalances for the entire dark halo, but also for the large four dark halo patches separately and compared these to our reference QS area. The entire area covered by the dark halo is relatively flux balanced. The relative imbalance derived is 4\% corresponding to 0.6\,G. The reference QS area shows a very similar value of 6\% corresponding to 0.8\,G of absolute flux imbalance. 
The individual dark halo patches exhibit different strengths of imbalance and even different signs, which can be found in Table~\ref{table}. THe patch DH1 partially overlaps with a negative polarity plage region at its innermost part, while the other dark halo patches have flux imbalances with a positive sign. The variation of these values might be due to limited statistics, as the areas covered by the individual dark halo patches are relatively small.

\section{Discussion}
\subsection{Comparison of dark halos to CHs and coronal voids}\label{S:discussion_comparison}
In addition to coronal dark halos, CHs and coronal voids are structures that also exhibit reduced emission in the EUV. Coronal holes show lower emission and hence an increasing difference to their surroundings towards higher coronal temperatures. In contrast, at low coronal temperatures the dark halo studied here generally displays a uniformly reduced emission below the QS level, while at higher temperatures its emission is significantly stronger than in the QS and decreases with distance from the centre of the AR. Thus, the emission from the dark halo patch DH1 in the 193\,\AA\ and 211\,\AA\ \aia\ channels consistently remains above or at the QS level. At these higher temperatures an equatorial CH becomes distinctly visible in the coronal emission from the dark halo patch DH3, where two clear dips in emission correlate with the location of the CH. This creates a noticeable contrast between dark halos and CHs at higher temperatures, making higher-temperature channels useful for distinguishing between the two.

Both, the dark halo and coronal voids \citep{Noelke2023} show a strong reduction in emission at 1\,MK compared to the QS. Most of the individual dark halo patches have a $\langle$|\blos|$\rangle$ that is slightly lower than that of the QS (89\%$-$94\% in DHs 2$-$4). The value of 115\% for DH1 is because it is cut off by the \hrt\ FOV (see Sect.~\ref{S:averageBLOS} and Fig.~\ref{blos_in_individual_dark_halos}). In contrast, such a difference is more marked for coronal voids, where the $\langle$|\blos|$\rangle$ is 76\% of the value in QS \citep{Noelke2023}.
However, we see a significant difference between $\langle$|\blos|$\rangle$ inside the dark halo and in the bright areas surrounding the dark halo patches (see Sect.~\ref{S:BLOS_distance}).
 
In addition to the 2021 November 5 Solar Orbiter observations, we further analysed SDO/HMI observations of 2018 April 22nd (see Appendix~\ref{appendixA}) studied by \cite{Lezzi2023}. The authors used a different technique to compare the magnetic field and derived the unsigned magnetic field, $B_{unsigned}$, which is the mean of absolute values of only those pixels above the noise level. By applying their technique and using a noise level threshold of 8\,G, we derived $B_{unsigned}=16.1$\,G for the QS and a $B_{unsigned}$ value of 15.6\,G. This value for the dark halo is slightly lower than the 17.0\,G reported by \cite{Lezzi2023}. The corresponding unsigned magnetic flux densities, $\langle$|\blos|$\rangle$, calculated using Eq.~\ref{Eq.fluxdensity} are 5.0\,G and 6.1\,G for the dark halo and QS, respectively.
Qualitatively, the result is the same as for the \hrt\ data: The $\langle$|\blos|$\rangle$ within the dark halo is, on average, weaker than in the QS. This difference is with 1.1\,G clearly stronger in the \hmi\ dataset, while the difference in the \hrt\ dataset is 0.6\,G or less (excluding DH1). It is also worth noting that, in the 2018 \hmi\ dataset, the $\langle$|\blos|$\rangle$ inside the dark halo remains weaker than in the QS at all distances from the AR centre, even at the inner boundary.

Dark halos typically show a systematic trend of the unsigned magnetic flux density dropping with radial distance from the AR, i.e from their boundaries closest to the AR centre to their outer boundaries (see Sect.~\ref{S:BLOS_distance} and Appendix~\ref{appendixA}). 
Close to the boundary to the AR the magnetic field is significantly stronger than in the QS. Then it drops to field strengths below the QS, and hence to values that compare well to coronal voids. In CHs and coronal voids the magnetic elements are more uniformly distributed than in the dark halos.

The dark halo patches show flux imbalances ranging from 10\% to 37\% (see Table~\ref{table} and Sect.~\ref{S:flux_imbalance}). Equatorial CHs on the other hand typically show much stronger flux imbalances of at least 50\% \citep{Wiegelmann2004,Hofmeister_2017}. The patch DH3, which overlaps with the equatorial CH, shows no significant deviation from the other dark halo patches.

\subsection{Nature of dark halos}\label{S:discussion_nature}
We confirm the finding of \cite{wang2011} that in the dark halos on average the $\langle$|\blos|$\rangle$ is slightly weaker than in the QS. However, our detailed analysis shows a more complex picture. The $\langle$|\blos|$\rangle$ drops with distance from the AR centre, and only in the outer parts of the dark halos the $\langle$|\blos|$\rangle$ is weaker than the QS. Still, at the same distance from the AR centre the $\langle$|\blos|$\rangle$ in the dark halo is weaker than in the bright areas between them (see Sect.~\ref{S:BLOS_distance})

The consistently lower value of $\langle$|\blos|$\rangle$ in the dark halo compared to the brighter regions between them could explain why the dark halos show reduced emission by direct comparison. A weaker magnetic field is indicative of less heating of coronal plasma and hence a lower intensity from plasma around 1\,MK. In the outer parts of the dark halo where the magnetic field is weaker than in the QS reduced coronal heating, similar to coronal voids in the QS \citep{Noelke2023}, could be a mechanism contributing to the reduced emission. However, while the magnetic field drops with distance from the AR centre, the emission seen in the \aia\ 171\,\AA\ and \euv\ channels is almost constant with distance. Moreover, because the $\langle$|\blos|$\rangle$ in the inner part of the dark halo is stronger than in the QS, one would naively expect the \aia\ 171\,\AA\ and \euv\ emission to be stronger in the inner part of the dark halo than in the QS, in contrast to our observations. 
Two possible scenarios that might explain this are described below.

Firstly, the heating of the corona depends not only on the magnetic field, but also on horizontal flows, that together set the vertical Poynting flux and hence the energy flux into the corona. Smaller horizontal flows near the AR could compensate for an increasing magnetic flux and keep the heating rate stable throughout the dark halo. This would in turn lead to the uniformly reduced intensity seen in the \aia\ 171\,\AA\ and \euv\ channels. Future research could explore this possibility further by investigating the horizontal flows in the vicinity of ARs.

Secondly, the emission from cooler plasma ($\leq$1\,MK), as seen in the structures in the enhanced image of the \aia\ 171\,\AA\ channel (see Fig.~\ref{Enhanced_images}), is organised in short loops. Within the dark halo fewer of these loops are visible. This is in direct contrast to plasma at higher temperatures ($\geq$1.6\,MK) probed by the \aia\ 193\,\AA\ and 211\,\AA\ channels, where the emission is organised in longer loops that often spread radially away from the AR. Interestingly, the emission observed in these passbands probing hotter plasma exhibits a similar radial dependence to that of the magnetic field. These longer loops predominantly connect to the stronger magnetic patches within the dark halo or to the brighter islands within the dark halo structure. Hot long-ranging loops have previously also been observed by \cite{singh2021}. Their model explains the difference between the hotter and cooler plasma seen in the different passbands with strong magnetic pressure exerted onto the small-scale loops that are then confined to stable heights below 5\,Mm.

\section{Conclusions}
We studied the coronal dark halo around the AR NOAA 12893. Specifically, we analysed the coronal emission at different temperatures and the photospheric magnetic field as a function of distance to the AR centre.

We report a radial profile in emission for the dark halo that varies with distance for higher coronal temperatures ($\geq$1.6\,MK) and shows reduced emission throughout the dark halo at temperatures $\leq$1\,MK (Sect.~\ref{S:coronal_emission}). Similarly, we detected variations in the unsigned magnetic flux density, $\langle$|\blos|$\rangle$, with distance (Sect.~\ref{S:magnetic_field}). These radial dependences of the emission and magnetic field imply that studies should not solely rely on mean values (e.g. averages over entire dark halos), as done in earlier studies.

The dataset further allowed us to analyse an equatorial CH adjacent to the AR (see Sects.~\ref{S:dark_halos_and_CH}, ~\ref{S:intensities_individualpatches_radialdistance}, and \ref{S:discussion_comparison}). Dark halos and CHs can clearly be distinguished via their emission at different coronal temperatures because CHs show a strong contrast above 1\,MK, whereas dark halos are only visible below these temperatures. This was also reported by \cite{Lezzi2023}.

In Sects.~\ref{S:introduction} and \ref{S:discussion_nature} we discuss existing models explaining the dark halos and propose an additional explanation: 
The source regions of hotter emission ($\geq1.6$\,MK; Sect.~\ref{S:intensities_individualpatches_radialdistance}) with visible loops over and reaching inside the dark halo (Sect.~\ref{S:long_loops}) are likely less dominated by the local magnetic field of the dark halo, but rather by the large-scale magnetic field of the AR. The dark halo shows a clear reduction of $\langle$|\blos|$\rangle$ compared to its brighter surroundings (Sect.~\ref{S:BLOS_distance}). At its outer boundaries it falls below the QS level. We suggest that reduced coronal heating due to reduced magnetic flux density might locally contribute to the dark appearance of the dark halos.

Future work could include an analysis of horizontal flows around ARs to test whether reduced flows within a dark halo near an AR lead to a smaller Poynting flux, resulting in lower heating, and thus reduced coronal emission. This will provide information on whether the reduced emission closer to the AR centre (where the $\langle$|\blos|$\rangle$ is stronger than in the QS) is caused by the suppressed horizontal flows. 
Obtaining new observations that include coronal spectroscopic measurements, for example from the SPICE instrument \citep{SPICEinstrument_2020}, would be immensely beneficial. These data would allow us to identify potential outflows by examining Doppler shifts within a dark halo. This would help us  either to identify clear boundaries of dark halos, CHs, and outflow areas or to reveal where these phenomena converge, thus allowing the further characterisation of this region. Studying more dark halos would also enable us to gather statistical data on the correlation between the radial distance to an AR and the $\langle$|\blos|$\rangle$ or the coronal emission.

\begin{acknowledgements}
Solar Orbiter is a space mission of international collaboration between ESA and NASA, operated by ESA. We are grateful to the ESA SOC and MOC teams for their support. The German contribution to SO/PHI is funded by the BMWi through DLR and by MPG central funds. The Spanish contribution is funded by AEI/MCIN/10.13039/501100011033/ and European Union “NextGenerationEU”/PRTR” (RTI2018-096886-C5, PID2021-125325OB-C5, PCI2022-135009-2, PCI2022-135029-2) and ERDF “A way of making Europe”; “Center of Excellence Severo Ochoa” awards to IAA-CSIC (SEV-2017-0709, CEX2021-001131-S); and a Ramón y Cajal fellowship awarded to David Suarez-Orozco. The French contribution is funded by CNES. 
The EUI instrument was built by CSL, IAS, MPS, MSSL/UCL, PMOD/WRC, ROB, LCF/IO with funding from the Belgian Federal Science Policy Office (BELSPO/PRODEX PEA 4000134088, 4000112292, 4000136424, and 4000134474); the Centre National d’Etudes Spatiales (CNES); the UK Space Agency (UKSA); the Bundesministerium für Wirtschaft und Energie (BMWi) through the Deutsches Zentrum für Luft- und Raumfahrt (DLR); and the Swiss Space Office (SSO). 
This work was supported by the International Max Planck Research School (IMPRS) for Solar System Science at the University of G\" ottingen and at TU Braunschweig. 
This project has received funding by the European Research Council (ERC) under the European Union's Horizon EUROPE research and innovation programme (grant agreements No. 101097844 — project WINSUN - and No. 101039844 --- ORIGIN). Views and opinions expressed are however those of the author(s) only and do not necessarily reflect those of the European Union or the European Research Council. Neither the European Union nor the granting authority can be held responsible for them. 
This research used version 4.1.5 (10.5281/zenodo.7850372) of the SunPy open source software package \citep{sunpy_community2020}.
\end{acknowledgements}

\begin{appendix}
\nolinenumbers
\section{SDO 2018 April 22 dataset}\label{appendixA}

\begin{figure}
\centering
    \resizebox{\hsize}{!}{\includegraphics{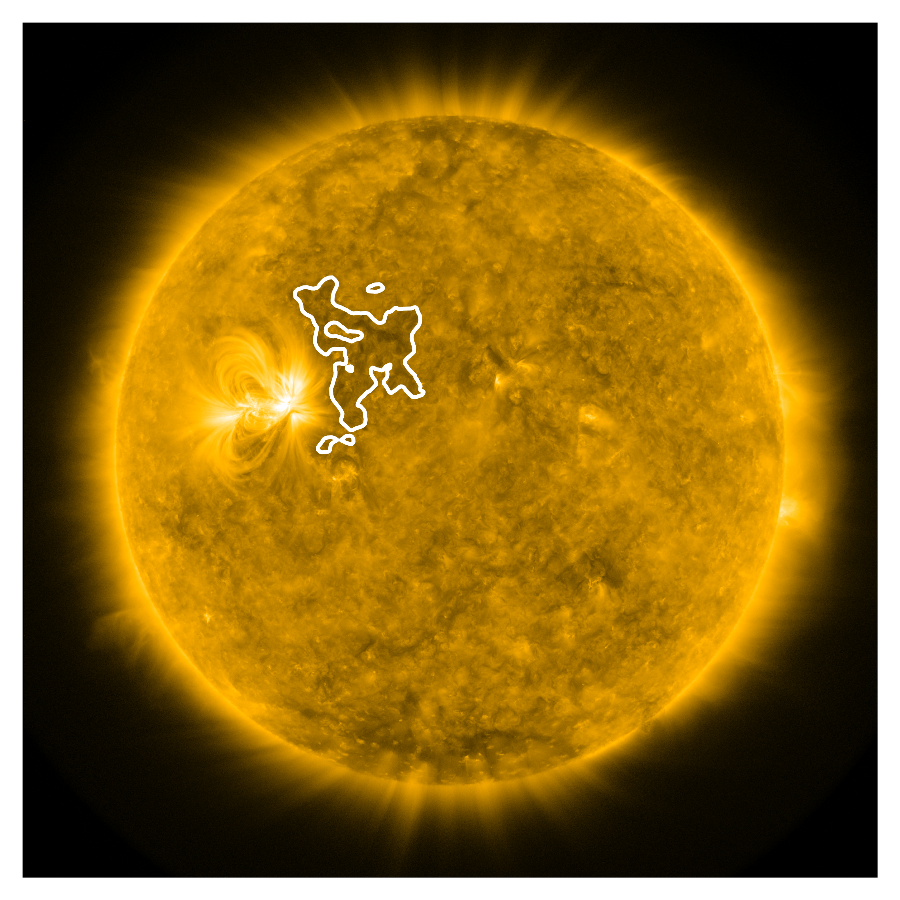}}
     \caption{SDO/AIA 171\,\AA\ observation of 2018 April 22 19:24. The dark halo adjacent to AR NOAA 12706 is outlined in white.}
    \label{overview_AIA_2018-04-22}
\end{figure}

\begin{figure}
\centering
    \resizebox{\hsize}{!}{\includegraphics{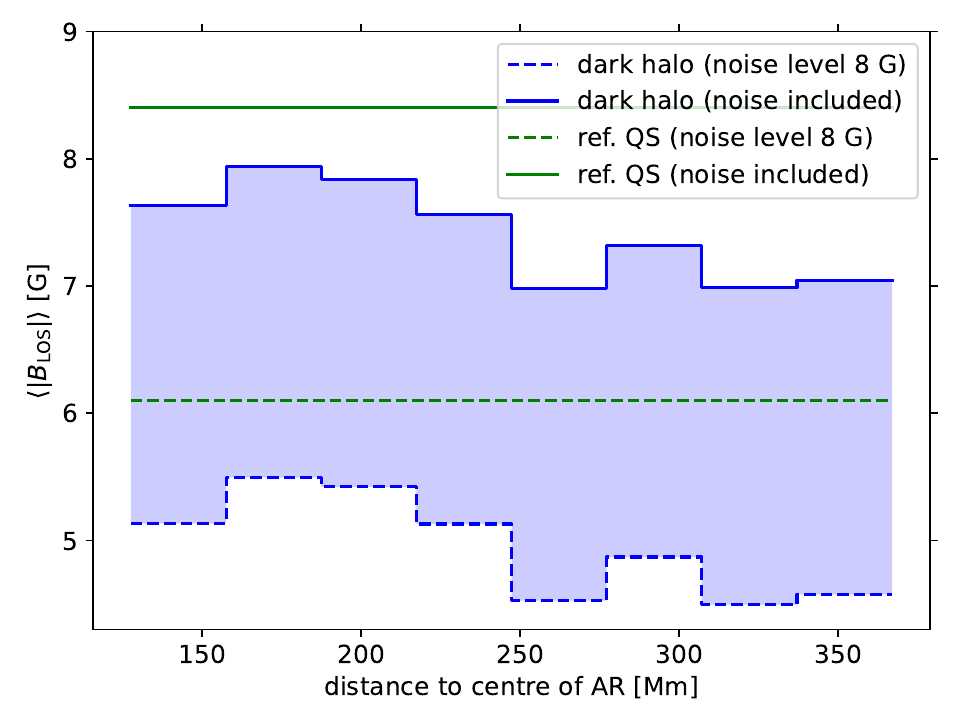}}
    \caption{$\langle$|\blos|$\rangle$ in dependence of distance to the AR centre for the 2018 April 22 SDO observations. We show the $\langle$|\blos|$\rangle$ inside the dark halo in blue calculated in two ways. The dashed line indicates the treatment where all pixels below the noise level have been set to zero, and the solid line displays the case where pixels below the noise levels have also been included. The shaded area indicates the range of these two calculations giving the maximum uncertainty range for the real unsigned magnetic flux density. The $\langle$|\blos|$\rangle$ for the QS is given in green.}
    \label{SDO2018_blos_distanceAR}
\end{figure}

\begin{figure}
\centering
    \resizebox{\hsize}{!}{\includegraphics{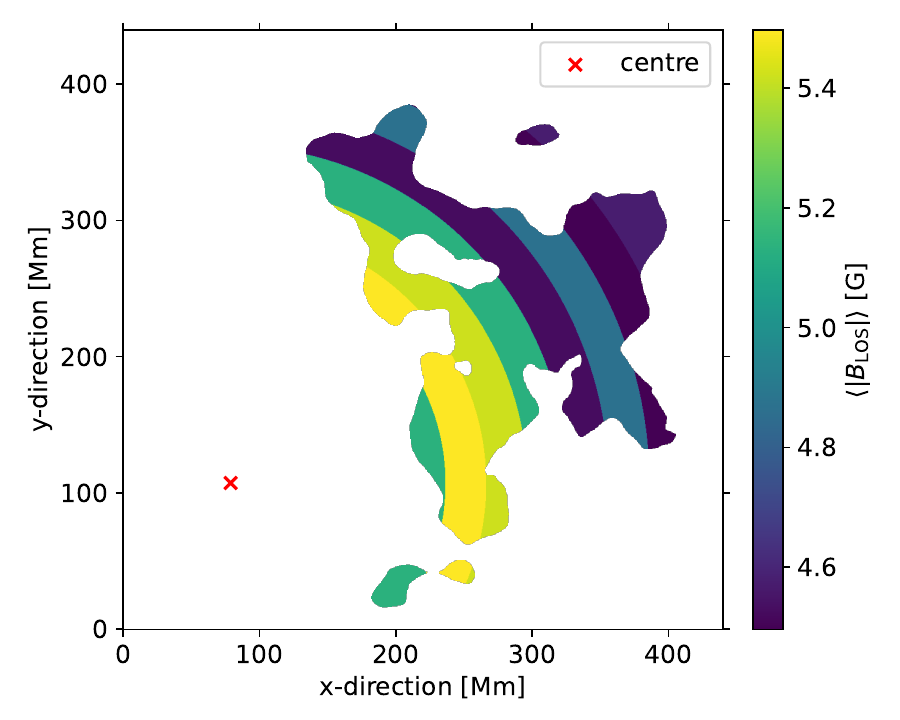}}
     \caption{ $\langle$|\blos|$\rangle$ inside concentric rings around the AR given as a function of distance to the AR centre for the SDO observations taken on 2018 April 22 19:24.}
    \label{BLOS_rings_SDO2018}
\end{figure}

\setcounter{figure}{0}
\renewcommand{\thefigure}{B\arabic{figure}}
\begin{figure*}
\centering
    \includegraphics[width=\textwidth]{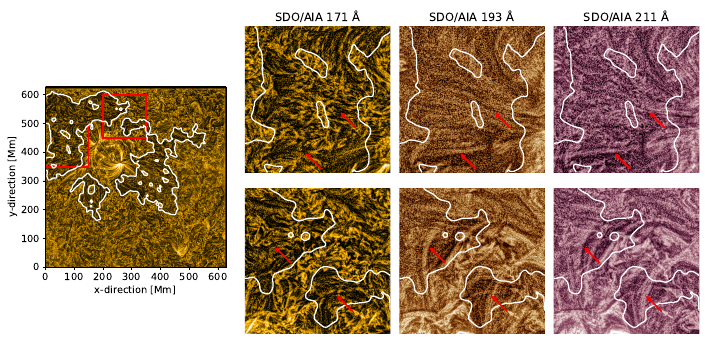}
     \caption{Different loop structures highlighted. The regions of interest are indicated by red squares on the enhanced 171\,\AA\ image (left panel), and the corresponding areas are shown for the 171\,\AA, 193\,\AA, and 211\,\AA\ images (right panels). The red arrows point to long-ranging loops present in the 193\,\AA\ and 211\,\AA\ channels, while absent in the cooler 171\,\AA\ one.}
    \label{different_loop_structures}
\end{figure*}

We studied the same SDO dataset from 2018 April 22 19:24 (with the HMI magnetogram from 19:24:24, and the three AIA channels at 171\,\AA, 193\,\AA, and 211\,\AA\ from 19:24:09, 19:24:04, and 19:24:09, respectively) as \cite{Lezzi2023}.
It contains AR NOAA 12706 and a dark halo adjacent to it in the west (see Fig.~\ref{overview_AIA_2018-04-22}). We aim to analyse the photospheric magnetic field of this dark halo and compare the results to the dark halos analysed in Sect.~\ref{S:BLOS_distance} and by \cite{Lezzi2023}. For this purpose we produced level 1.5 data for the 171\,\AA\ observation and additionally normalised the intensity to DN/s by dividing the image by the exposure time. The co-temporal \hmi\ LOS magnetogram was re-projected to the \aia\ image using the SunPy reproject package.

In order to reproduce the dark halo contours in \cite{Lezzi2023} we tested different filters and filter settings for convolving the \aia\ 171\,\AA\ image from which smooth iso-contours of the dark halo can be obtained. A uniform filter with a size of 40 pixels yields the closest result to theirs. We first convolved the 171\,\AA\ image with this uniform filter and then applied a 55\% mean disc intensity threshold to identify the dark halo next to the AR NOAA 12706. These contours are very similar to the ones that can be obtained by convolving the image with a Gaussian filter and then applying an intensity threshold of 75\% of the mean QS intensity at disc centre (as has been done in Sect.~\ref{S:dark_halo_boundaries}). We also selected the same reference QS area in the north-east of the solar disc which has a size of $250\arcsec\times337\arcsec$.

We applied the same noise level of 8\,G as used by \cite{Lezzi2023} to derive the unsigned magnetic field, $B_{unsigned}$, by averaging the absolute values in the pixels above this noise level. We obtained $B_{unsigned}=15.6$\,G in the dark halo and $B_{unsigned}=16.1$\,G in the QS. The QS value compares well to that presented in \cite{Lezzi2023}; however, the value for the dark halo is 1.4\,G lower than theirs. 

To better compare these values with the 2021 November 8 Solar Orbiter dataset, we calculated the corresponding unsigned magnetic flux densities, $\langle$|\blos|$\rangle$ (according to Eq.~\ref{Eq.fluxdensity}). We applied the same noise level of 8\,G and derived a lower limit for the unsigned magnetic flux density by setting pixels below the noise level to zero. This way we obtained an unsigned magnetic flux density of 5.0\,G for the dark halo and the higher value of 6.1\,G for the QS. 
This result is consistent with our previous findings for the 2021 November 5 Solar Orbiter observations in the sense that the $\langle$|\blos|$\rangle$ in the DH is somewhat lower than in the QS. From Fig.~\ref{SDO2018_blos_distanceAR} it follows that this is the case at every distance from the centre of the neighbouring AR, irrespectively of whether we consider the upper or the lower limit of the flux density.

In the same way as for the 2021 November 5 Solar Orbiter observation, we calculated the $\langle$|\blos|$\rangle$ inside concentric rings around the AR. We derived the centre of the AR by computing the geometric centre of magnetic elements with field strengths above 100\,G within an area of $\sim$175\,Mm (400~pixels) around the clearly visible leading sunspot. As in Sect.~\ref{S:BLOS_distance} we calculated the $\langle$|\blos|$\rangle$ inside each concentric ring with a width of 30\,Mm starting at $\sim$130\,Mm (290~pixel). Like in Sect.~\ref{S:BLOS_distance} we derived $\langle$|\blos|$\rangle$ in two ways: by considering only signals above the noise level are solar (i.e. setting all signals hidden in the noise to zero), and by assuming that pixels below the noise level show signal (i.e. assuming that the noise is zero). This is shown in Fig.~\ref{BLOS_rings_SDO2018}.

Although only a relatively weak decrease in the unsigned magnetic flux density from the inner to the outer boundaries of this dark halo is observed, the magnetic field tends to decrease with distance from the AR. This $\langle$|\blos|$\rangle$--distance dependence is shown in Fig.~\ref{SDO2018_blos_distanceAR}. 
In the ring closest to the AR centre the dark halo has a value of 5.1\,G (considering the noise corrected values), which rises to 5.5\,G inside the next ring, but then starts to decrease showing some variation in this trend. At the outer edge of the dark halo $\langle$|\blos|$\rangle$ has reached 4.6\,G. This general trend agrees with that already observed in the 2021 November 5th Solar Orbiter observations (see Fig.~\ref{blos_distanceAR}). However, the difference between the dark halo and the QS is more pronounced in this dataset. The QS is with 6.1\,G well above the $\langle$|\blos|$\rangle$ inside the dark halo at all distances from the AR.

\section{Visibility of long-ranging loops in different channels}\label{appendixB}
The appearance of long-ranging loops spanning over the dark halo differs between the \aia\ channels. While these loops are clearly visible in the enhanced 193\,\AA\ images, they appear less prominent in the 211\,\AA\ images probably because the lower signal-to-noise and stronger background emission reduce the contrast relative to the surrounding corona. In the 171\,\AA\ channel, probing cooler plasma, such long loops are not present.

Fig.~\ref{different_loop_structures} provides a close-up view of two selected regions (indicated in the right panel) where these differences are most apparent. The regions of interest are indicated by red squares on the enhanced 171\,\AA\ image (left panel), and the corresponding areas are shown for the 171\,\AA, 193\,\AA, and 211\,\AA\ images (right panels). This comparison shows that the same long loops can be identified in both the 193\,\AA\ and the 211\,\AA\ channels, though fainter in 211\,\AA. They are absent, however, in 171\,\AA.

\end{appendix}

\end{document}